\documentclass[preprint,5p,times,twocolumn]{elsarticle}

\usepackage{amssymb}
\usepackage{graphicx}
\usepackage{epstopdf}
\usepackage{booktabs}
\usepackage{amsmath}
\usepackage{enumitem}
\usepackage{multirow}
\usepackage[linesnumbered,ruled,vlined]{algorithm2e}

\journal{Arxiv Preprint}

\begin{document}

\begin{frontmatter}




 \title{Leveraging Large Language Models for Enhanced Digital Twin Modeling:\\ Trends, Methods, and Challenges}

 \author[aff1]{Linyao Yang\corref{cor1}}
 \affiliation[aff1]{organization={Research Center for Data Hub and Security, Zhejiang Lab},
             city={Hangzhou},
             postcode={311121},
             country={China}}
 \cortext[cor1]{corresponding author}
 \ead{yangly@zhejianglab.org}

 \author[aff1]{Shi Luo}

 \author[aff2]{Xi Cheng}
 \affiliation[aff2]{organization={Research Center for Intelligent Manufacturing Computing, Zhejiang Lab},
             city={Hangzhou},
             postcode={311121},
             country={China}}

 \author[aff1]{Lei Yu}

\begin{abstract}


Digital twin technology is a transformative innovation driving the digital transformation and intelligent optimization of manufacturing systems. By integrating real-time data with computational models, digital twins enable continuous monitoring, simulation, prediction, and optimization, effectively bridging the gap between the physical and digital worlds. Recent advancements in communication, computing, and control technologies have accelerated the development and adoption of digital twins across various industries. However, significant challenges remain, including limited data for accurate system modeling, inefficiencies in system analysis, and a lack of explainability in the interactions between physical and digital systems. The rise of large language models (LLMs) offers new avenues to address these challenges. LLMs have shown exceptional capabilities across diverse domains, exhibiting strong generalization and emergent abilities that hold great potential for enhancing digital twins. This paper provides a comprehensive review of recent developments in LLMs and their applications to digital twin modeling. We propose a unified description-prediction-prescription framework to integrate digital twin modeling technologies and introduce a structured taxonomy to categorize LLM functionalities in these contexts. For each stage of application, we summarize the methodologies, identify key challenges, and explore potential future directions. To demonstrate the effectiveness of LLM-enhanced digital twins, we present an LLM-enhanced enterprise digital twin system, which enables automatic modeling and optimization of an enterprise. Finally, we discuss future opportunities and challenges in advancing LLM-enhanced digital twins, offering valuable insights for researchers and practitioners in related fields.

\end{abstract}



\end{frontmatter}


\section{Introduction}

With the rapid advancement and widespread adoption of the Internet of Things (IoT) \cite{HANSEN2021362}, 5G/6G communication technologies, and artificial intelligence (AI) \cite{LENG2024349}, efficient and accurate monitoring, optimization, and decision-making for manufacturing systems have become increasingly achievable. Digital twin technology is a transformative innovation that enables the monitoring, simulation, prediction, and optimization of physical systems by creating digital replicas of physical entities or processes. By integrating these digital replicas with real-time data streams, digital twins provide decision-makers with a comprehensive understanding of their operations \cite{10570372}, empowering them to predict system behavior, optimize performance, and make data-driven decisions to enhance efficiency and productivity. However, conventional digital twin frameworks often overlook critical social factors, such as economic, ethical, and ecological considerations, and are still heavily reliant on manual efforts for analyzing and optimizing physical systems. In contrast, parallel systems—a similar but more automated control paradigm for complex systems—facilitate the automated analysis and adaptive optimization of physical systems through virtual experiments, enabling the identification of optimal solutions within the digital domain.

\begin{figure}[!htp]
  \centering
  \includegraphics[width=0.95\linewidth]{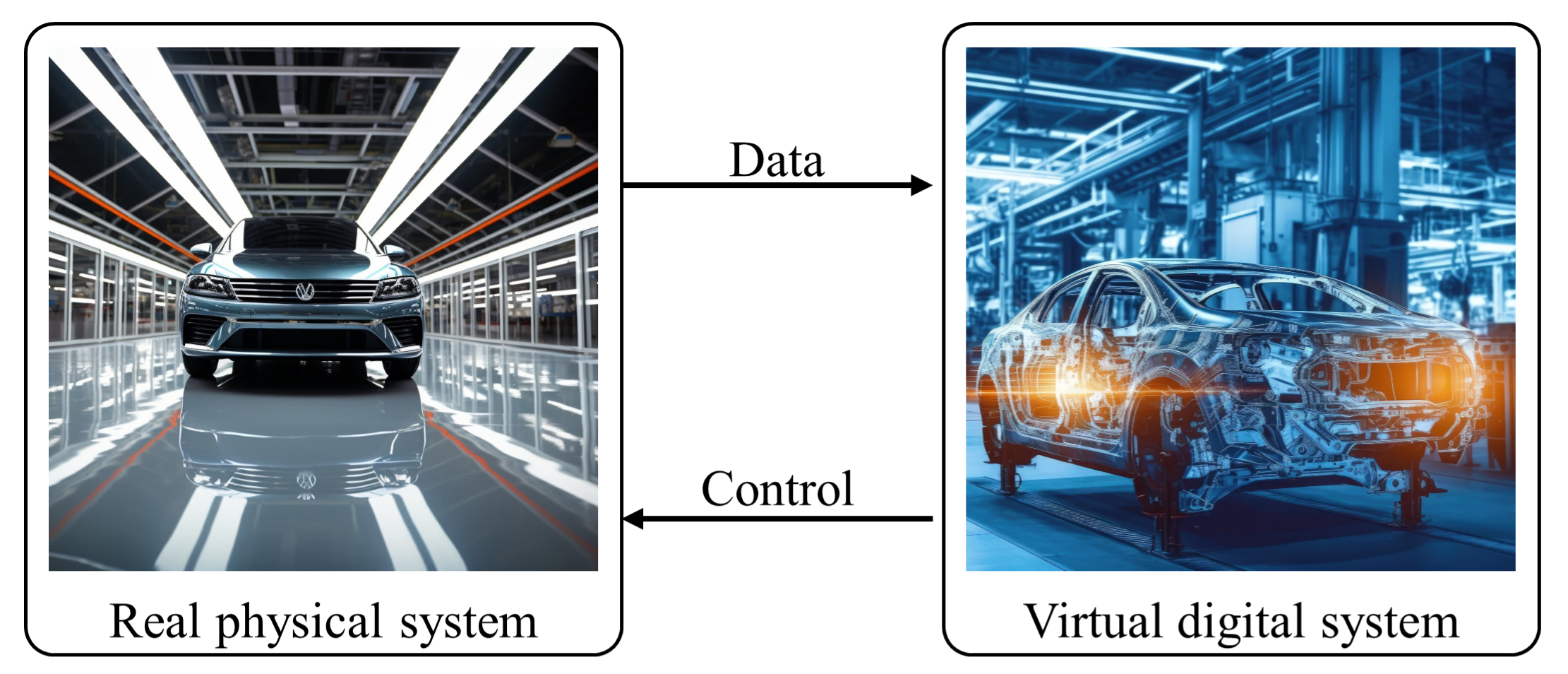}
  \caption{Foundamental architecture of digital twins.}
\label{fig:frameworks}
\end{figure}


The concept of digital twins originated from a presentation titled "Conceptual Ideal for Product Lifecycle Management," delivered by Michael Grieves at the University of Michigan in 2003 \cite{grieves2005product}. This concept evolved into the foundational architecture of digital twins, which includes three key components: the real physical system, the virtual digital system, and the communication link between them, as shown in Figure~\ref{fig:frameworks}. The virtual digital system acts as a digital replica of the physical system, serving as a virtual laboratory where experiments that cannot be performed on physical systems can be conducted. It not only reflects the state of the physical system but also allows for the optimization of control strategies through computational analysis. Furthermore, recent advancements in AI and intelligent control techniques have enhanced the digital twin framework by integrating diverse analytical and control technologies, improving its ability to perform automatic analysis and control for physical systems \cite{WANG20241045,KARKARIA2024322}. For example, Wang et al. \cite{WANG20241045} proposed a digital twin-based field-of-view adjustment method that optimizes the layout for photogrammetry cameras using a 3D digital twin. Similarly, Karkaria et al. \cite{KARKARIA2024322} developed a machine learning-based digital twin framework for real-time model predictive control in directed-energy deposition processes. These developments suggest that digital twins are evolving toward greater predictability and control.

In recent years, digital twins have rapidly advanced, with numerous methodologies proposed \cite{TAO2022372,7974888,8699108,XIE2024475} and widespread applications emerging in fields such as manufacturing, transportation, and healthcare \cite{WANG2021165,LAI202376,10260246,9114349}. Most existing digital twin frameworks can be incorporated into a unified description-prediction-prescription model. In the description stage, digital systems are developed to represent the mechanisms, states, and dynamics of the physical system, based on data collected from the physical system. In the prediction stage, data from both physical and digital systems are analyzed to derive optimal strategies for improving the system. Finally, in the prescription stage, the learned strategies are implemented in the physical system to guide its evolution toward an improved state.

These advancements have significantly increased the level of automation and intelligence in various industrial domains. However, several challenges remain in modeling and implementing these systems. First, digital system development often relies on real-world data or manually crafted rules to create software-defined simulation models \cite{8226990}. However, real-world data for specific physical systems is often insufficient or inaccessible due to privacy concerns. Additionally, creating accurate simulation models is time-consuming and labor-intensive. Second, human interactions within traditional digital systems are often implicit and inefficient \cite{ZHOU2020124779}, making it difficult to model complex group behaviors effectively. Third, the design and implementation of data analysis processes require significant human effort and specialized expertise, which hinders the efficiency of developing optimization strategies. Finally, current control mechanisms often involve either recommending solutions to managers \cite{9788029} or directly deploying them in physical systems, but these approaches lack sufficient interactivity and interpretability, limiting their wider applicability.

Language, as the primary medium of human expression, plays a crucial role in modeling human knowledge and describing the laws of the physical world. Language models, which predict the generative likelihood of word sequences, enable machines to comprehend and generate human language, facilitating the acquisition of human knowledge. Recent breakthroughs in large language models (LLMs) \cite{GPT-3,nijkamp2022codegen} have positioned them as powerful world simulators, capable of encapsulating vast amounts of real-world knowledge and demonstrating remarkable emergent capabilities \cite{wei2022emergent}. Leveraging these strengths, LLMs have been successfully applied in various aspects of system modeling and optimization, including information retrieval \cite{zhu2023large,mao-etal-2023-large,ziems-2023-large}, data processing \cite{3458754,BERT,10121440}, system simulation \cite{xia2024llm,wang2024does,3643385,rafailov2024direct,aher2023using,gao2023s}, experimental design \cite{nijkamp2022codegen,3543957}, and strategy generation \cite{zhang2021greaselm,10160591,friedman2023leveraging}. These advancements open new possibilities for modeling digital twins, a field that remains largely underexplored.

To address this gap, we present a systematic review of recent advancements in LLMs and their applications in system modeling and optimization. This review is organized into three categories—descriptive modeling, predictive modeling, and prescriptive modeling—aligned with the digital twin modeling framework. We identify specific modeling tasks that can benefit from LLMs, such as data collection, data analysis, and explainable control, and critically evaluate the existing methods. To demonstrate the potential of LLM-enhanced digital twins, we develop an enterprise digital twin system powered by LLMs. Additionally, we explore the challenges and highlight promising research directions for further advancing LLM-enhanced digital twins.

\subsection{Related Review Works}

Digital twins have rapidly emerged as a dynamic area of research in recent years, leading to an increasing number of publications, including several review articles. To clarify the novelty of this paper, we provide a brief overview of existing reviews, highlighting their limitations.

Barbara Rita et al. \cite{8901113} reviewed the state-of-the-art definitions, key characteristics, and applications of digital twins. Semeraro et al. \cite{SEMERARO2021103469} provided an overview of ongoing research and technical challenges in the design and development of digital twin systems, with a focus on their core components, features, and interaction challenges. Lim et al. \cite{lim2020state} summarized the concepts and techniques associated with digital twins throughout their lifecycle. However, these reviews primarily focus on definitions, concepts, and the historical evolution of digital twins, offering limited analysis of the modeling techniques employed in digital twin systems. In contrast, this paper shifts the focus from foundational definitions to addressing the core tasks involved in digital twin modeling, with a particular emphasis on exploring innovative solutions enabled by LLMs.

Tao et al. \cite{TAO2022372} conducted a systematic review of digital twin modeling, examining the various models from the perspective of application fields and enabling technologies. Liu et al. \cite{LIU2021346} provided an in-depth review of the key enabling technologies for digital twins. Mihai et al. \cite{9899718} reviewed essential enabling technologies, challenges, and future prospects for digital twins, focusing on design challenges and limitations across different industries. Qin et al. \cite{10570372} explored the role of machine learning and deep learning techniques in enhancing the efficiency of digital twin networks. Miao et al. \cite{10057176} systematically reviewed recent advancements in training models using virtual data, categorizing approaches to solving physical system problems with data derived from digital systems. Although these reviews cover diverse modeling techniques, significant challenges remain in advancing these technologies. Moreover, the rapid progress of emerging technologies, such as LLMs, has not yet been comprehensively examined in the context of digital twins.

Other reviews have focused primarily on specific applications of digital twins. Leng et al. \cite{LENG2021119} introduced an innovative framework to examine the applications of digital twins in smart manufacturing systems (SMS). Their review addresses definitions, frameworks, major design steps, new blueprint models, enabling technologies, design cases, and future research directions for digital twin-based SMS. Wu et al. \cite{9429703} offered a detailed discussion on definitions, key technologies, and typical application scenarios of digital twin networks. Lin et al. \cite{lin2024human} conducted an extensive review of human digital twin (HDT) technologies, highlighting their applications across healthcare, industry, and daily life. Yang et al. \cite{yang2019digital} provided a comprehensive comparison of digital twins and parallel systems, analyzing their concepts, key technologies, and applications.

\subsection{Purpose of This Paper}

The emergence of LLMs opens new opportunities to address the typical challenges encountered in the modeling of digital twins. To foster further research on LLM-enhanced digital twin modeling, this paper presents a systematic review of LLM applications in related tasks, highlighting their potential to advance digital twin modeling. Compared to existing surveys, the main contributions of this paper are as follows:

\textbf{Unified framework.} This paper unifies the key processes of digital twin modeling into a description-prediction-prescription framework, identifying three stages where LLMs can provide enhancements.

\textbf{Task identification and review.} The paper identifies fundamental tasks within each stage that can benefit from LLMs, systematically reviewing recent advancements in LLM-enhanced approaches. This review provides valuable insights for optimizing modeling processes. To the best of our knowledge, this is the first comprehensive review focusing on LLM-enhanced digital twin modeling.

\textbf{Case analysis.} A case study is presented to demonstrate the effectiveness of LLM-enhanced digital twin modeling.

\textbf{Challenges and future directions.} The paper analyzes the challenges faced in current research and proposes several promising directions for future investigation.

The remainder of this paper is organized as follows:

Section \uppercase\expandafter{\romannumeral2} provides background information on digital twins, parallel systems, and LLMs. Section \uppercase\expandafter{\romannumeral3} presents the description-prediction-prescription framework, identifying key modeling tasks that can be enhanced by LLMs and analyzing existing methods in detail. Section \uppercase\expandafter{\romannumeral4} develops an LLM-enhanced enterprise digital twin system to demonstrate the effectiveness of LLMs in improving digital twin modeling. Section \uppercase\expandafter{\romannumeral5} discusses the challenges faced by current LLM-enhanced modeling methods and outlines potential future research directions. Section \uppercase\expandafter{\romannumeral6} concludes the paper.

\section{Backgrounds}

Digital twins begin with the creation of digital systems based on data collected from physical systems, which are then used to monitor and optimize the performance of the corresponding physical systems. Parallel systems, a related paradigm for managing and controlling complex systems, extend traditional digital twins to cyber-physical-social systems (CPSS) and enable more intelligent and automated control of physical systems through computational analysis and virtual-real interactions. However, current modeling methods for these systems face several challenges. LLMs, with their vast repository of world knowledge and remarkable emergent capabilities, offer new opportunities to enhance the performance and efficiency of various modeling tasks. This section provides an overview of key concepts and backgrounds related to digital twins, parallel systems, and LLMs.

\subsection{Digital Twins}

\begin{figure}[!htp]
  \centering
  \includegraphics[width=1.0\linewidth]{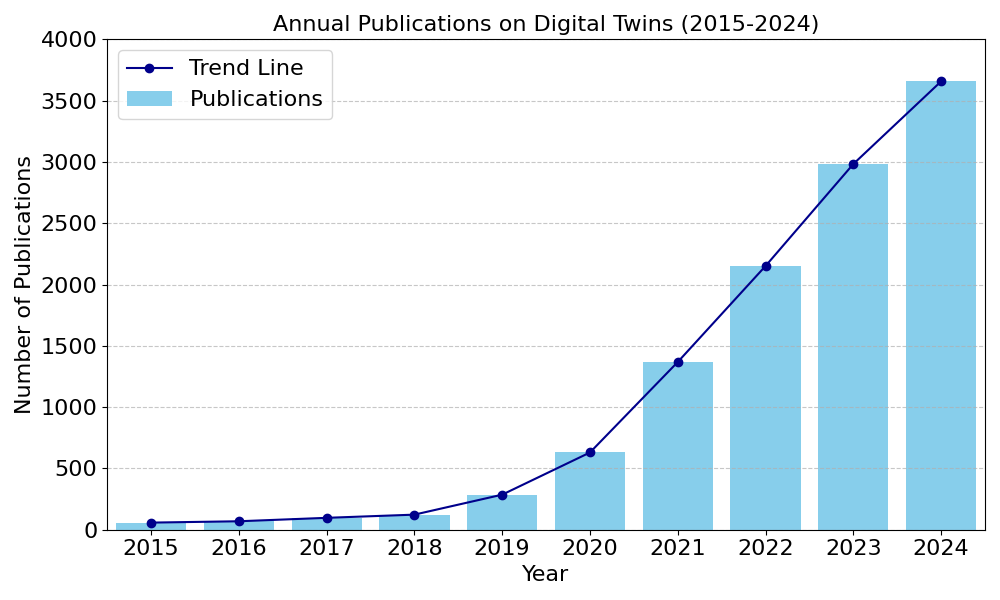}
  \caption{Number of papers published per year over the past ten years.}
\label{fig:pubs}
\end{figure}

Over the past decade, both interest and research in digital twins have experienced significant growth. To assess the development trends in this field, we conducted a search in the Web of Science database using "digital twin" as the keyword. Based on the retrieved articles, Figure~\ref{fig:pubs} illustrates the annual number of publications related to digital twins over the past ten years, showing a steady increase each year. This underscores the rapid growth and expanding research interest in digital twins.

Although the concept of digital twins can be traced back to Grieves' report in 2003 \cite{grieves2005product}, the term "digital twin" was formally introduced in 2010 by the National Aeronautics and Space Administration (NASA) \cite{shafto2010draft}. The core idea of digital twins is to create a digital representation of a physical entity to evaluate, optimize, and manage that entity through interactions between the digital model and its physical counterpart. Despite the simplicity of this concept, numerous conceptual and reference models for digital twins have been proposed. For example, Grieves \cite{grieves2014digital} introduced a three-dimensional conceptual model consisting of three components: physical products, virtual products, and their data and information connections. Zheng et al. \cite{zheng2019application} proposed an application framework for digital twins in product lifecycle management, comprising three key components: the physical space, the virtual space, and the information processing layer. Tao et al. \cite{8477101} developed an extended five-dimensional architecture that models digital twin systems in five dimensions: the physical entity, virtual model, connection, digital twin data, and service.

The construction of digital twin systems encompasses a variety of technologies, including information, communication, and computing technologies, which significantly drive the advancement of digital twins. Mihai et al. \cite{9899718} identified six key enabling technologies for digital twins: machine learning, cloud, fog, and edge computing, the Internet of Things (IoT), cyber-physical systems, virtual reality (VR) and augmented reality (AR), and modeling methodologies. As these technologies continue to develop, more intelligent and autonomous digital twin systems are emerging, enhancing the automatic modeling and optimization of digital twin processes. Chen et al. \cite{CHEN2023581} summarized the role of machine learning in digital twins, revealing that machine learning-powered digital twins can significantly expedite the development of predictive maintenance systems. He et al. \cite{HE2019221} proposed a data-driven digital twin system for automatic process applications, integrating virtual modeling, process monitoring, diagnosis, and optimized control into a cooperative architecture. As a result, contemporary digital twin systems are progressively expanding into the realm of the description-prediction-prescription framework.

\begin{figure}[!htp]
  \centering
  \includegraphics[width=1.0\linewidth]{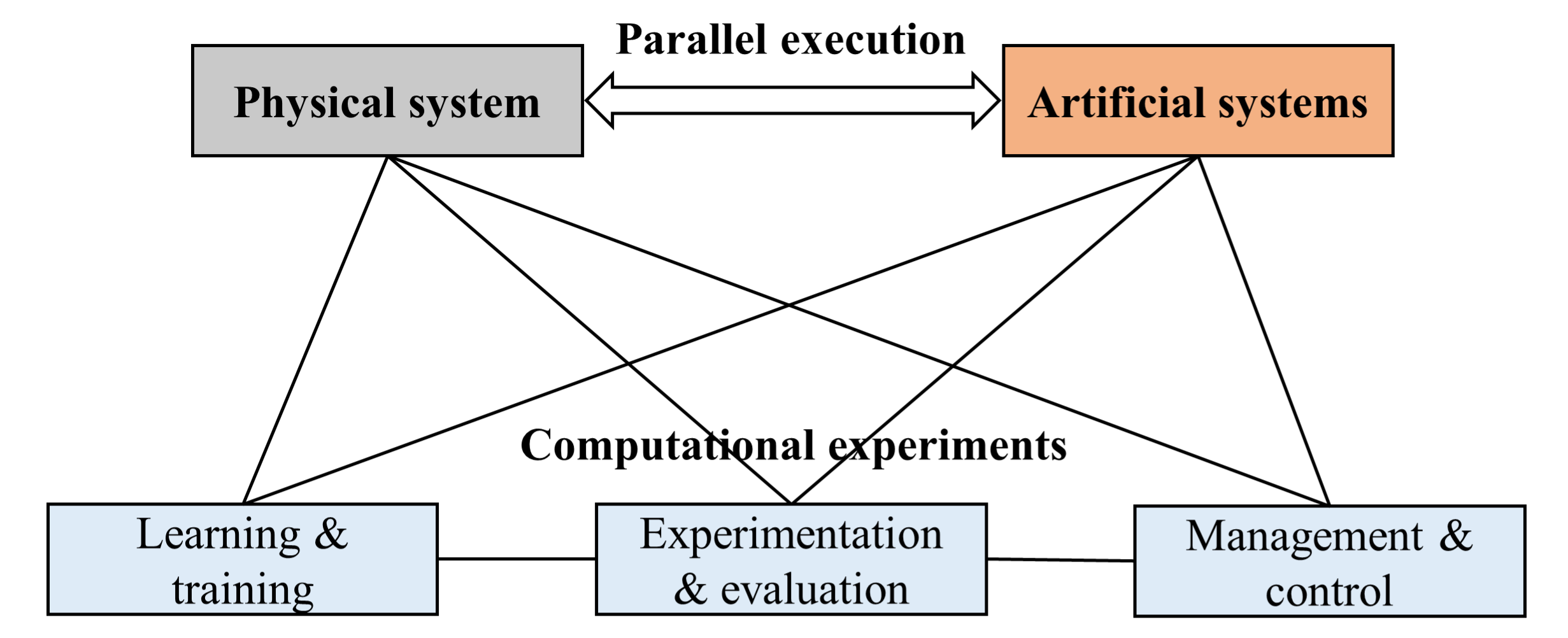}
  \caption{Architecture of the ACP approach.}
\label{fig:ACP}
\end{figure}

\subsection{Parallel Systems}

Parallel systems \cite{PS-1} aim to use experimentable artificial systems to learn optimal strategies for various scenarios and optimize physical systems through virtual-real interactions between artificial and physical systems. This approach is similar to digital twins but offers more automation in control. The fundamental method for developing parallel systems is the ACP approach—artificial systems, computational experiments, and parallel execution, as illustrated in Figure~\ref{fig:ACP}.

The first step in developing a parallel system is to construct artificial systems, which serve as digital testbeds for experiments that would be impractical in physical systems. Most existing studies use agent-based modeling to simulate various elements based on data collected from the physical system or human-crafted rules. For example, Yang et al. \cite{9051726} employed a decision tree model to analyze questionnaire data and developed agents to simulate pedestrian evacuation choices. Ye et al. \cite{8226990} proposed a general cognitive architecture to encompass all aspects of decision-making in artificial societies. However, real-world data is often insufficient, with some scenarios lacking data, and security or privacy concerns further limit data availability. To address these issues, data augmentation methods \cite{wang2017parallel,8699108} have been proposed to generate synthetic data for more effective agent modeling. Additionally, reinforcement learning \cite{8370297,9016408} is widely used to train agents in open environments. Despite these advancements, modeling artificial systems still faces challenges, particularly with inefficient data collection and the extensive human effort required for designing data learning methods.

In the computational experiments step, various scenarios with different parameters and control strategies are established within artificial systems. Simulating these scenarios generates extensive artificial data \cite{8451919}, which can then be analyzed to derive optimal control strategies for specific situations \cite{PI-1}, utilizing advanced big data and artificial intelligence techniques. For instance, Li et al. \cite{8451919} proposed a systematic method for constructing large-scale artificial scenarios and created a new virtual dataset for traffic vision research. Chang et al. \cite{10530101} developed a framework for interaction scenario engineering in vehicles, integrating intelligent systems for transport automation. Zhu et al. \cite{8824092} assessed the impact of different control strategies on traffic by analyzing generated artificial data. Despite these advancements, the methods used to learn valuable control knowledge from computational experiments often require substantial human effort to design experimental setups and develop programs, which reduces the efficiency of these experiments and hinders the widespread application of parallel systems.

The parallel execution step continuously optimizes both artificial and physical systems through their virtual-real interactions. Typically, the optimal control strategy for the artificial scenario most similar to the physical system is either implemented directly to the physical system or recommended to a manager to ensure safety and reliability. The physical system then provides feedback to the artificial systems, which automatically update their states and control strategies. Zhang et al. \cite{8851791} introduced a feature adaptation module to address data distribution mismatches between synthetic and real data, making the object instance segmentation model trained on artificial data more applicable to real-world data. Jin et al. \cite{9788029} developed a recommendation system for strategic urban traffic management, using a regional agent dispatcher to assign agents as needed for "operation-on-demand." Additionally, Jin et al. \cite{9005386} proposed an end-to-end recommendation framework for urban traffic management and control, mimicking and enhancing the behaviors of professional signal control engineers. However, many existing parallel execution methods lack explainability, limiting their application in safety-sensitive systems. Furthermore, these methods often require complex interfaces for device manipulation within physical systems, demanding significant human effort.

Therefore, both digital twins and parallel systems face modeling challenges in terms of descriptive, predictive, and prescriptive modeling.

\subsection{Large Language Models}

LLMs are a type of language model pre-trained on extensive corpora, enabling them to capture the nuances of language. After pre-training, LLMs undergo fine-tuning \cite{LoRA} with human-annotated data tailored to specific downstream tasks. Additionally, they are optimized using reinforcement learning with human feedback \cite{RLHF}, which helps them follow human instructions while minimizing the generation of toxic, biased, or harmful content. Recently, numerous LLMs \cite{GPT-3, nijkamp2022codegen, CLIP} have been developed and successfully applied across various domains. With the increase in model size and training data, LLMs exhibit emergent abilities \cite{wei2022emergent}. To reflect this broader scope, we also consider foundational models with multimodal capabilities, which typically align multimodal representations with textual semantics.

The most notable emergent abilities of LLMs include \cite{10417790}: (1) Zero-shot learning, where LLMs can accurately perform certain tasks without task-specific training data, often outperforming traditional supervised models \cite{bang-etal-2023-multitask}. (2) In-context learning, which allows LLMs to learn from context without additional training \cite{dong2022survey}, enhancing their flexibility and adaptability. (3) Step-by-step reasoning, where LLMs can integrate intermediate reasoning steps into their responses using chain-of-thought prompting strategies \cite{COT}, improving performance on complex reasoning tasks. (4) Instruction following, which allows LLMs to follow human instructions to perform previously unseen tasks without explicit examples \cite{wei2021finetuned}. (5) Human value alignment, where LLMs are fine-tuned to generate responses that align with human values, using reinforcement learning from human feedback. (6) Tools manipulation, where LLMs are trained to manipulate third-party tools to perform complex tasks \cite{qin2023tool}. Furthermore, LLMs implicitly capture extensive world knowledge within their vast parameters.

These capabilities enable LLMs to be used as world simulators for developing digital mirror models of physical systems, achieving significant success in various system optimization tasks. For example, DriveDreamer \cite{zhao2024drivedreamer} uses an LLM to generate user-defined driving videos for training driving perception methods. Panacea \cite{wen2024panacea} leverages pretrained foundation models to create world models for simulating autonomous driving. Additionally, methods such as Copilot use LLMs to accelerate the control of robotic arms in physical systems. However, these approaches still face challenges in understanding knowledge across different domains and require improvements in system design, feedback optimization, and control reliability.

Leveraging these capabilities, LLMs are increasingly utilized as world simulators to develop digital mirror models of physical systems, achieving notable success across various system optimization tasks. For example, DriveDreamer \cite{zhao2024drivedreamer} employs an LLM to generate user-defined driving videos for training driving perception methods, while Panacea \cite{wen2024panacea} utilizes pretrained foundation models to create driving world models for simulating autonomous driving. Furthermore, LLMs have been effectively applied across a range of system modeling and optimization domains, including data collection \cite{zhu2023large,mao-etal-2023-large,ziems-2023-large}, data analysis \cite{3458754,BERT,10121440}, system modeling \cite{xia2024llm,wang2024does,3643385,rafailov2024direct,aher2023using,gao2023s}, experimental design and code generation \cite{pallagani2024prospects,nijkamp2022codegen,3543957}, as well as strategy generation and recommendation \cite{zhang2021greaselm,10160591,friedman2023leveraging}. These advancements highlight the potential of LLMs to revolutionize digital twin modeling, an area that remains underexplored.

To address this gap, we present a systematic review of recent advancements in LLMs and their applications in fields related to digital twin modeling.

\begin{figure}[!htp]
  \centering
  \includegraphics[width=1.0\linewidth]{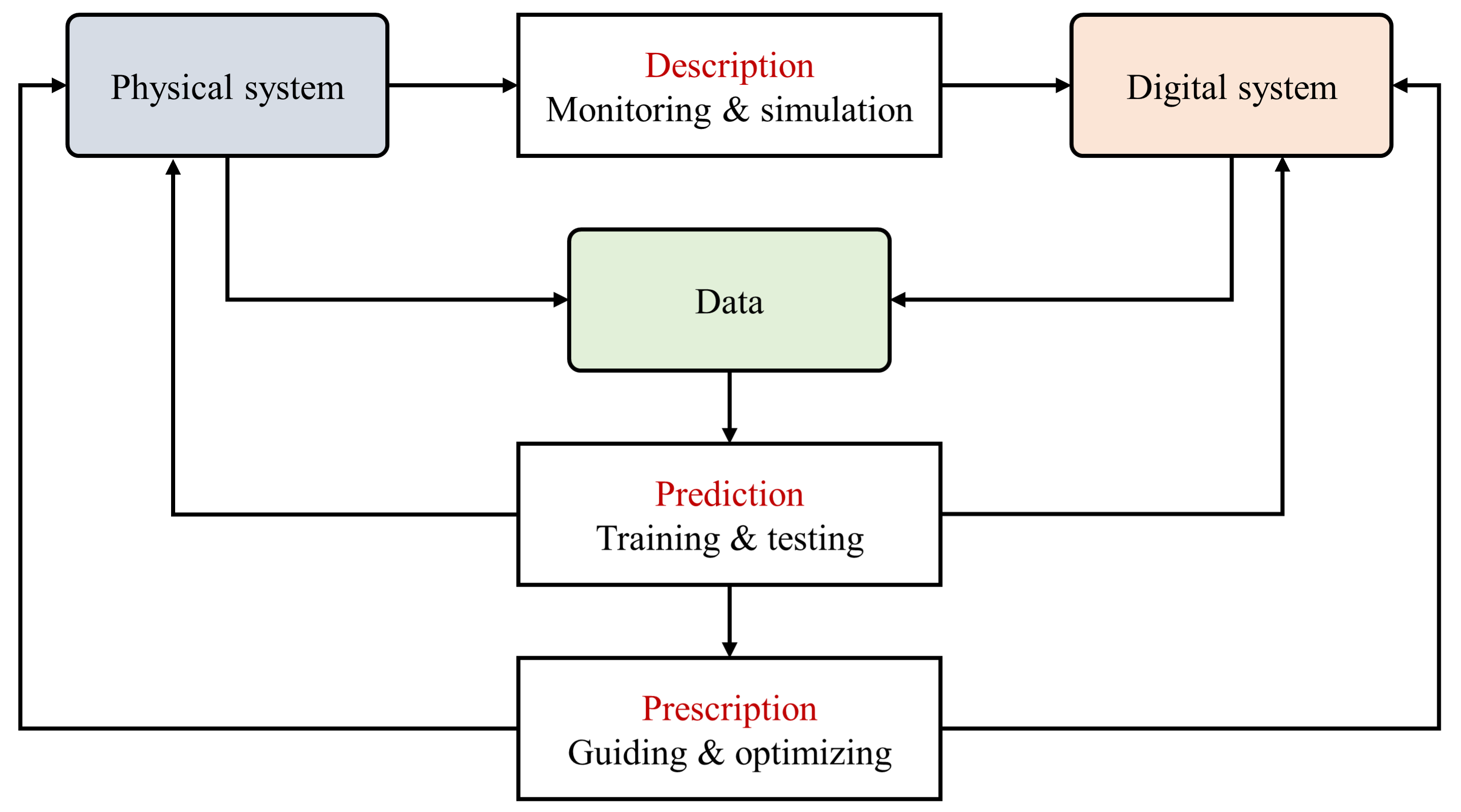}
  \caption{Architecture of the description-prediction-prescription framework.}
\label{fig:DPP}
\end{figure}

\section{Large Language Model-enhanced Digital Twin Modeling}

In this section, we introduce the description-prediction-prescription framework, which unifies the modeling processes of modern digital twins. We then categorize the roles of LLMs within this framework and identify key tasks in digital twin modeling. Finally, we systematically review existing LLM-based methods that can be adapted to enhance these processes.

\begin{figure*}[!htp]
  \centering
  \includegraphics[width=0.9\linewidth]{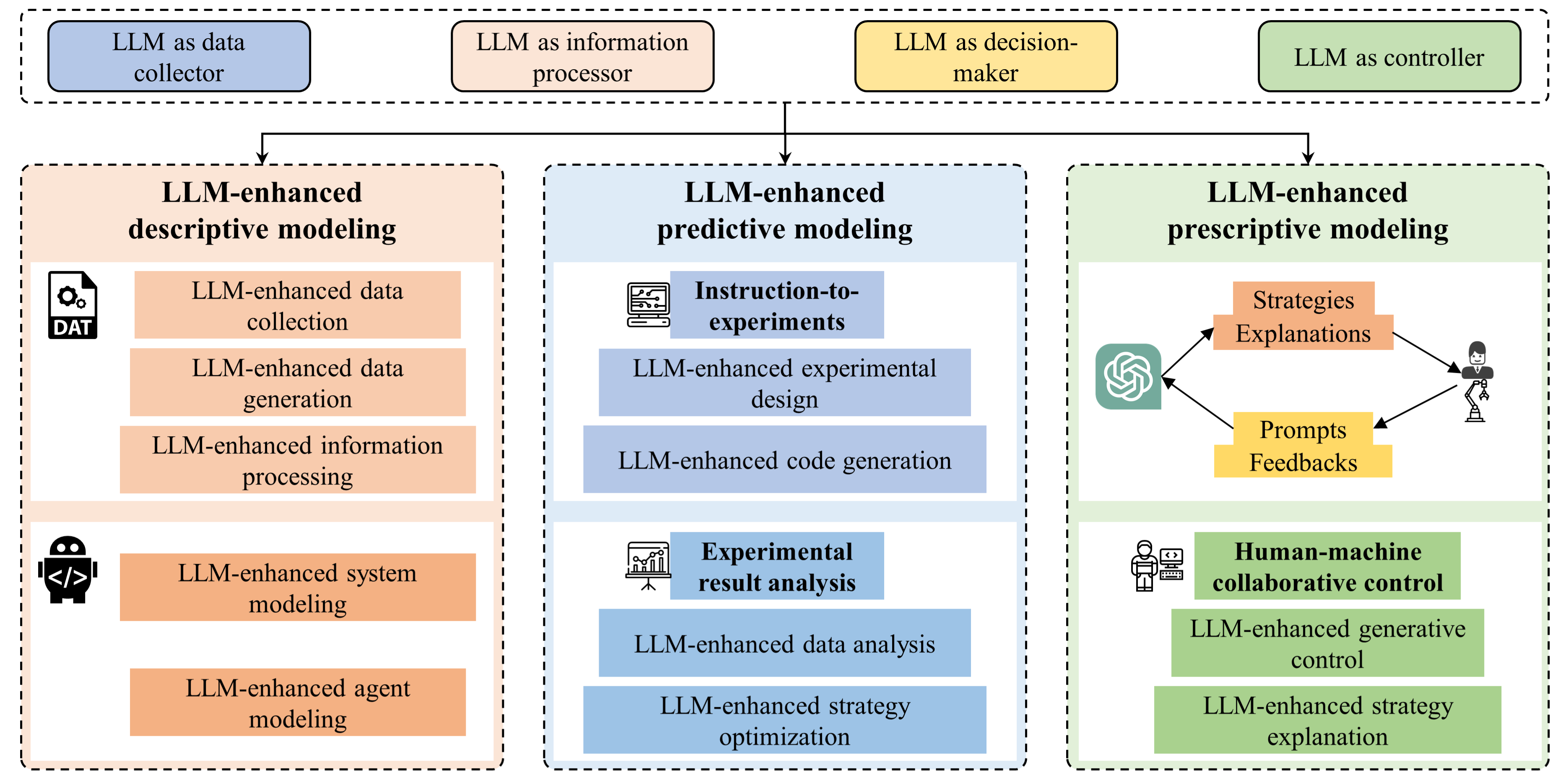}
  \caption{Tasks that can be enhanced by LLMs, where LLMs play different roles in enhancing descriptive, predictive, prescriptive modeling.}
\label{fig:LLM-roles}
\end{figure*}

\subsection{Framework}

The description-prediction-prescription framework, illustrated in Figure~\ref{fig:DPP}, provides a comprehensive structure for the modeling processes of digital twins. In this framework, the modeling of digital twins begins with the construction of one or more digital models through the collection of data and knowledge from the physical system, followed by analysis of the acquired information. The descriptive modeling phase encompasses tasks such as data acquisition, data analysis, and simulation modeling. In the predictive modeling stage, digital models are used to monitor or simulate the evolution of the physical system under varying conditions. This stage enables the design of experiments to evaluate the effects of different control strategies, with optimization methods applied to identify the most effective strategies. Finally, in the prescriptive stage, the identified control strategies are either directly implemented in the physical system or recommended to managers. These strategies induce state changes in the physical system, which in turn update the digital system via virtual-real interactions.

LLM-enhanced digital twin modeling leverages the advanced information processing, reasoning, and generative capabilities of LLMs to support the development and operation of digital twins. With their extensive emergent abilities, LLMs can enhance various aspects of digital twin modeling:

\textbf{Data acquisition and processing.} LLMs can improve the extraction of valuable information from physical systems, facilitating the construction of high-quality digital models. They can also generate synthetic data to address real-world data scarcity using their sophisticated generative capabilities.

\textbf{Simulation and reward definition.} LLMs can act as agents to simulate elements of the physical system, assisting in defining reward functions for agent-based learning. This capability is particularly useful in scenarios requiring detailed modeling of complex systems.

\textbf{Predictive modeling.} LLMs can convert text-based simulation instructions into digital models, automating experimental execution through their instruction-following and code generation abilities. They can also analyze and visualize experimental results, leveraging their reasoning capabilities to derive optimal strategies.

\textbf{Generative control.} Once an optimal strategy is determined, LLMs can implement it by converting it into executable code or interfacing with external tools. They can also generate natural language explanations for the strategy, enhancing human understanding and ensuring safety and reliability.

\textbf{Feedback-driven optimization.} LLM-enhanced digital systems can be iteratively improved based on feedback from the physical system using fine-tuning algorithms. This continuous optimization strengthens the alignment between digital and physical systems.
 
In summary, LLMs offer transformative potential for digital twin modeling by addressing challenges in data acquisition, simulation, prediction, and prescription. Their ability to seamlessly integrate reasoning, generation, and interaction makes them invaluable for advancing digital twin systems.

Based on their roles within the framework, the functions of LLMs in LLM-enhanced digital twins can be categorized into four main areas:

\begin{itemize}[itemsep=0pt, topsep=0pt]
\item \textbf{Data collector.} The development of digital systems relies on extensive data about various aspects of the physical system, which is often difficult to acquire, imbalanced, or incomplete. To address these challenges, LLMs can function as data collectors by retrieving relevant information from diverse sources or generating synthetic data to augment existing datasets, thus ensuring the availability of high-quality data for digital system modeling.

\item \textbf{Information processor.} Modeling digital twins often involves handling multisource and multimodal data, which can be complex and difficult to interpret. Pretrained on large-scale corpora and aligned across multiple languages and modalities, LLMs excel at representing data in semantically rich vector spaces and extracting actionable insights from raw data. As information processors, LLMs can generate task-specific feature representations and distill valuable insights, simplifying subsequent learning tasks.

\item \textbf{Decision-maker.} Leveraging their advanced capabilities in generation, reasoning, and instruction-following, LLMs can serve as decision-makers in several critical tasks: 1) simulating physical systems as self-organizing agents; 2) designing and executing experiments based on user instructions; and 3) analyzing experimental results to derive optimal strategies.

\item \textbf{Controller.} Controlling physical systems requires seamless interaction with machines and humans, a task traditionally managed through manually crafted interfaces. With their tool manipulation and reasoning abilities, LLMs can learn to operate machines based on provided documentation. Additionally, they can explain their actions, offering insights into the rationale behind specific decisions, thereby enhancing system transparency.
\end{itemize}

\begin{figure*}[!htp]
  \centering
  \includegraphics[width=0.9\linewidth]{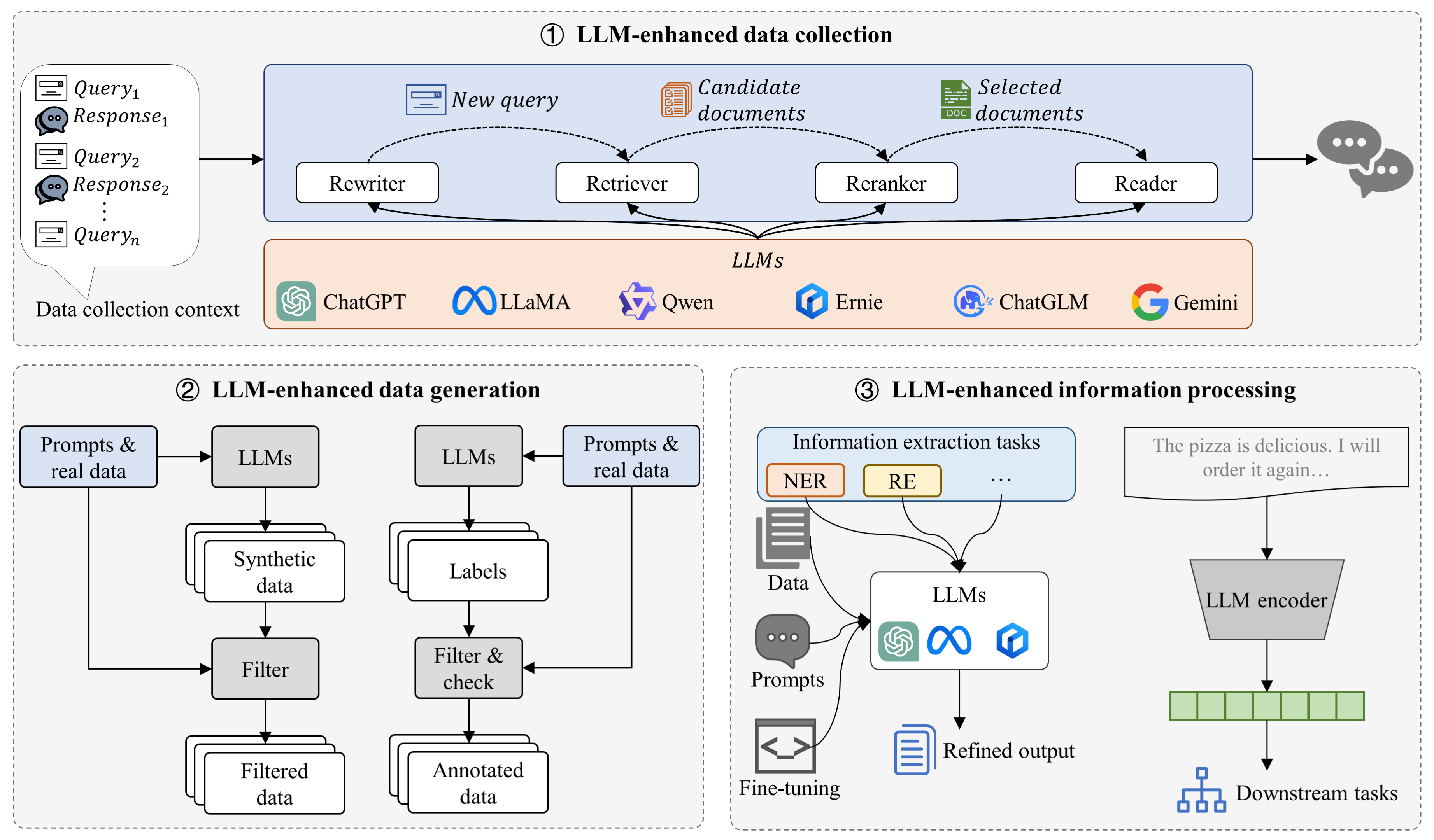}
  \caption{LLM-enhanced data engineering framework for the modeling of digital systems.}
\label{fig:LLM4Data}
\end{figure*}

The following section discusses how LLMs can be utilized to perform various tasks in digital twin modeling. Their roles and corresponding tasks are illustrated in Figure~\ref{fig:LLM-roles}.

\subsection{LLM-enhanced Descriptive Modeling}

The traditional approach to constructing digital systems involves designing agent-based simulation models for various elements of the physical system based on prior knowledge or sensor data. However, real-world data often suffers from issues such as incompleteness and noise, which create significant challenges for agent modeling. Additionally, small, hand-crafted agent models are difficult to train due to limited supervision and human feedback, and they are challenging to scale across different elements and scenarios. Complex agents with extensive capabilities and intricate action spaces cannot be adequately represented by these small models. As a result, higher-quality data and more powerful agent models are essential for effective digital system modeling.

By pretraining on large-scale corpora, LLMs acquire extensive knowledge of data patterns and encode high-level semantics into vector spaces. Leveraging their robust representation capabilities and prior knowledge, LLMs can function both as data collectors and information processors. They enhance real-world data by retrieving additional relevant information or generating high-quality synthetic data, while also extracting clean and meaningful insights from noisy inputs. Furthermore, LLMs excel in language understanding, instruction-following, planning, and reasoning tasks, making them promising candidates for developing general-purpose agents. In summary, LLMs contribute to descriptive digital system modeling from two key perspectives: data engineering and agent modeling.

\subsubsection{LLM-enhanced data collection}

Traditional digital system modeling methods often rely on various sensors \cite{8824092,9374560} to collect data from physical systems, leading to significant redundancy and increased demands for data transmission. In contrast, LLMs possess the capability to interpret and compress multimodal sensing data, offering a novel approach to improving data management and operational efficiency through advanced compression and decompression techniques. For instance, Yang et al. \cite{Transcompressor} proposed an LLM-driven multimodal data compression framework for smart transportation, demonstrating the effectiveness of LLMs in reconstructing transportation sensor data. Similarly, Hong et al. \cite{hong2024llm} introduced an LLM-enabled digital twin networking framework that redefines intra-twin and inter-twin communication architectures. By leveraging LLMs for semantic-level communication and computation, this framework enhances both communication and computational efficiency. Additionally, Jiang et al. \cite{10757470} developed an LLM-integrated framework for digital twin networks that minimizes transmission delays by selectively retrieving relevant data.

Information retrieval systems are also valuable tools for gathering desired data from diverse sources, such as web pages and databases. Recently, efforts have been made to leverage the power of LLMs to improve the performance of information retrieval systems, reshaping the landscape of data collection. Consequently, LLMs can be employed to enhance data collection for digital system modeling. Specifically, LLMs can improve traditional information retrieval components, such as query rewriting, retrieval, and reranking \cite{zhu2023large}. For example, Query2doc \cite{Query2Doc} generates pseudo-documents through few-shot prompting of LLMs, which are then used to expand the query. LLMCS \cite{mao-etal-2023-large} leverages LLMs as text-based search intent interpreters to enhance conversational search. Nan et al. \cite{nan-etal-2023-enhancing} explored various prompt design strategies to use LLMs for converting texts into SQL queries.

By applying these techniques, LLMs can generate more accurate and contextually relevant query rewrites. They can also generate synonyms and related concepts based on their extensive knowledge, thereby broadening the coverage of queries to include a wider range of relevant data. LLMs can further assist in developing superior retriever models for collecting relevant data based on specific queries. Leveraging the strong text understanding and generation capabilities of LLMs offers a promising solution for filtering out redundant or noisy information from the collected data. Moreover, utilizing the fine-grained encoding and decoding abilities of LLMs enables more precise calculation of the semantic relevance between queries and documents \cite{BGE}.

LLMs have also been successfully applied to rerank documents, serving as a second-pass document filter in information retrieval. Some methods fine-tune LLMs to fully understand the reranking task, improving their ability to measure query-document relevance \cite{nogueira-etal-2020-document}. Other methods prompt LLMs to enhance document reranking directly in an unsupervised manner \cite{zhuang-etal-2023-open}. Additionally, researchers have explored expanding the scope of information retrieval systems beyond document ranking to answer generation by leveraging LLMs as readers, based on their extensive capabilities in understanding, extracting, and summarizing \cite{lazaridou2022internet}.

Therefore, by integrating LLMs into information retrieval systems, data required for modeling various physical elements can be efficiently collected from diverse sources, providing more comprehensive data support for agent-based modeling.

\subsubsection{LLM-enhanced data generation} 

Traditional modeling methods for digital systems often require substantial amounts of labeled data \cite{9051726}, but the process of data annotation is time-consuming and labor-intensive. Although some approaches \cite{8699108} utilize generative adversarial networks or simulation models to produce synthetic labeled data, these methods still face significant limitations, including the need for extensive domain expertise and challenges related to training instability. In contrast, LLMs provide a more flexible and efficient approach to data generation.

LLMs, having been pre-trained on large and diverse datasets, are capable of generating high-quality, contextually relevant virtual data across various domains with minimal additional training. For instance, Zhou et al. \cite{10700677} introduced an LLM-driven framework that combines LLMs with domain-specific generative models to synthesize general sensor data. Similarly, Meng et al. \cite{meng2022generating} prompted a unidirectional LLM to generate class-conditional texts, which were then used to fine-tune a bidirectional LLM, significantly improving its performance. InPairs \cite{bonifacio2022inpars} pairs documents with a collection of query-document pairs to prompt GPT-3 to generate relevant queries for each document. VL2NL \cite{VL2NL} uses LLMs to generate diverse and rich natural language datasets based on Vega-Lite specifications, streamlining the development of natural language interfaces for data visualization. GReaT \cite{borisov2022language} leverages an autoregressive generative LLM to generate synthetic yet highly realistic tabular data by conditioning on any subset of features, effectively modeling tabular data distributions. Additionally, LLMs are widely employed to generate images, videos, and other data formats using prompts \cite{CLIP}, substantially reducing the cost of generating diverse synthetic data for agent modeling.

Moreover, LLMs have made tremendous advancements in few-shot learning tasks, which can be utilized as low-cost labelers to train agent models. For example, Wang et al. \cite{wang-etal-2021-want-reduce} explored ways to use GPT-3 for annotating unlabeled data, significantly reducing the cost of labeling compared to human annotation while achieving comparable performance. Recently, LLM-annotated data has been widely adopted for training smaller models, providing an effective method for knowledge distillation from LLMs. Therefore, LLMs can be leveraged to augment real data by generating diverse synthetic data, offering valuable insights for digital system modeling.

\subsubsection{LLM-enhanced information processing} 

Existing studies employ various data analysis methods to extract useful information from collected data for agent learning \cite{9114349,8039015}, typically requiring extensive feature engineering, large labeled datasets, and task-specific models. As a result, these traditional approaches often face challenges related to scalability, adaptability, and the effective use of contextual information. In contrast, LLMs offer more robust, flexible, and context-aware data analysis capabilities.

On one hand, LLMs excel at extracting key information from large volumes of data due to their advanced comprehension and summarization capabilities. Recent LLMs have been enhanced with long-context processing abilities, enabling them to handle inputs containing millions of tokens. For instance, Liu et al. \cite{liu-text-summarize} proposed a novel document-level encoder based on BERT to capture document-level semantics and generate sentence-level representations. Agrawal et al. \cite{agrawal-etal-2022-large} demonstrated that LLMs can serve as effective clinical information extractors, achieving exceptional performance in zero-shot and few-shot information extraction from clinical texts. Similarly, Fei et al. \cite{bbaa110} integrated large-scale biomedical knowledge graphs into an LLM to extract structured information from biomedical texts. Additionally, Zhang et al. \cite{10547418} developed a multimodal LLM for multisensor image comprehension in remote sensing, unifying various multisensor remote sensing tasks within a single framework.

On the other hand, LLMs can generate high-quality embeddings that capture intricate patterns and contextual nuances in the data, thus improving the performance of downstream learning tasks. For example, Yang et al. \cite{IPEA} utilized pretrained word embeddings from LLMs to initialize name embeddings for entities, improving knowledge fusion performance across different knowledge graphs. Xu et al. \cite{xu-etal-2019-bert} post-trained BERT on domain-specific data to obtain better word representations for tasks like review reading comprehension and sentiment analysis. Furthermore, LLMs are adept at seamlessly integrating multimodal data, creating comprehensive representations by aligning the semantics of different modalities in shared embedding spaces \cite{CLIP}, which enhances the effectiveness of multimodal data analysis. LLMs can also generate code and manipulate tools to perform more customized data analysis tasks.

In conclusion, LLMs can function as powerful information processors to extract valuable insights from both collected and generated data for agent modeling, providing scalable and robust data analysis support.

\begin{figure*}[!htp]
  \centering
  \includegraphics[width=0.9\linewidth]{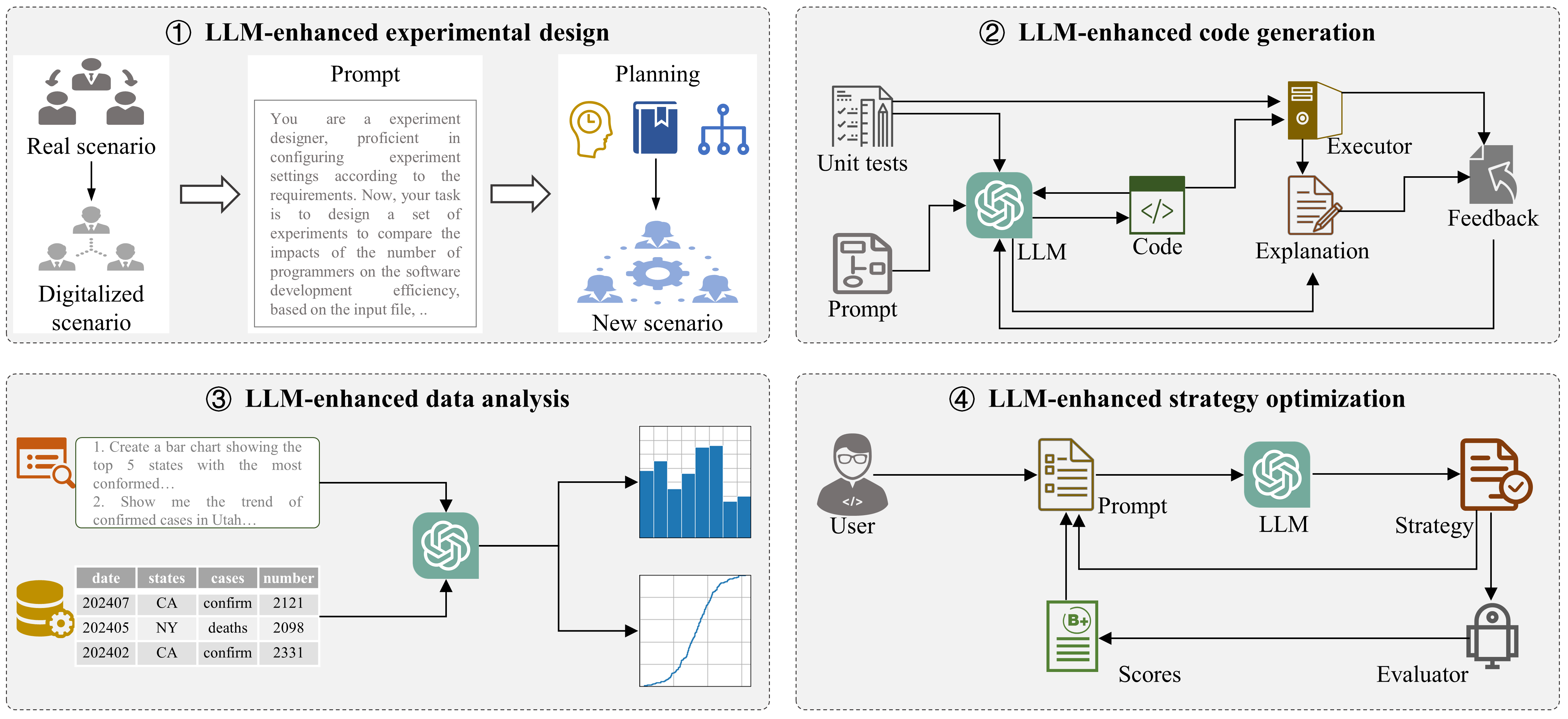}
  \caption{Aspects in predictive modeling that can be enhanced by LLMs.}
\label{fig:LLM4C}
\end{figure*}

The framework for LLM-enhanced data engineering in digital system modeling is illustrated in Fig.~\ref{fig:LLM4Data}. With the support of LLMs, higher-quality data can be acquired and analyzed, providing more valuable information for modeling various elements of the physical system. Based on the acquired information, LLMs can also be employed to enhance agent models, either by acting as agents themselves or by providing supervision for their learning.

\subsubsection{LLM-enhanced system modeling}

Existing industrial digital twins often rely on textual description documents to construct digital models, using various modeling languages such as JSON, Universal Scene Description (USD), and XML to develop different representations. However, designing high-fidelity digital models presents significant challenges, as it requires extensive manual effort and domain expertise to edit these models or their associated description documents. As the complexity and scale of the systems grow, manual editing becomes increasingly labor-intensive and error-prone. Therefore, innovative methods for automating scene generation are highly desirable.

Pretrained on autoregressive generation tasks, LLMs offer substantial potential for generating hierarchical description documents from textual prompts. Li et al. \cite{ChatTwin} introduced a conversational system that leverages GPT-4 to streamline the creation of comprehensive model description documents specifically for data center digital twins. Höllein et al. \cite{Text2Room} proposed a method for generating room-scale textured 3D meshes from text prompts, utilizing pretrained 2D text-to-image models to synthesize a sequence of images from various perspectives. Xia et al. \cite{10710900} applied LLMs to automate the parametrization of simulation models within digital twins, enabling dynamic interactions with simulations to explore parameter configurations and determine feasible settings for specific objectives. Similarly, Gebreab et al. \cite{10786996} presented a digital twin framework that uses LLMs to automate and accelerate critical stages of digital twin development, including 3D modeling.

The integration of LLMs into digital simulation model development highlights their transformative potential in automating and enhancing model generation processes. By leveraging their capabilities in natural language understanding, hierarchical structure generation, and multimodal synthesis, LLMs can significantly reduce reliance on manual effort and domain expertise. This not only accelerates the development of high-fidelity digital models but also minimizes errors associated with manual editing, especially as system complexity scales.

\subsubsection{LLM-enhanced agent modeling}

Agents are software-defined entities capable of simulating physical elements by processing desires, beliefs, intentions, and actions. Agent-based modeling is widely used in the construction of digital systems \cite{8226990}. However, most existing agent-based modeling approaches rely heavily on manually designed models, which are time-consuming, labor-intensive, and difficult to extend to different physical elements.

The development of LLMs offers a promising solution to the challenges faced by agent modeling in digital systems, and recent studies have made significant progress in this area. One approach to LLM-enhanced agent modeling involves directly leveraging LLMs to construct agents. Some studies use LLMs as the primary component for the brain or controller of agents, expanding their perceptual and action capabilities through strategies such as multimodal perception and tool manipulation. Specifically, LLM-based agents typically consist of three components: perception, brain, and action \cite{xi2023rise}. The perception module processes information from the external environment, the brain module stores crucial memories, information, and knowledge, and undertakes essential tasks such as information processing, decision-making, reasoning, and planning. Meanwhile, the action module executes tasks using tools and influences the environment. LLMs enhance the simulation of human behavior by generating human-like text, engaging in conversations, and performing complex tasks without requiring detailed step-by-step instructions. For instance, Wang et al. \cite{Twin-GPT} proposed an LLM-based digital twin creation approach that generates unique, personalized digital twins for different patients, preserving individual patient characteristics. Research has shown that with proper conditioning, LLMs can accurately emulate response distributions across various human subgroups \cite{argyle2023out}. Additionally, LLMs exhibit strong transferability and generalization to unseen tasks, making them easily customizable for different types of agents without the need for task-specific training.

Another approach to improving agent modeling is by providing agents with rewards or feedback from LLMs. With their powerful modeling capabilities and extensive world knowledge, LLMs can function as world model simulators, learning complex environmental dynamics with high fidelity \cite{cao2024survey}, thus facilitating agent learning. For example, ELLM \cite{3618754} rewards an agent for achieving goals suggested by a language model prompted with a description of the agent's current state, guiding agents toward human-meaningful and plausibly useful behaviors without requiring human intervention. Kwon et al. \cite{kwon2023reward} used an LLM as a proxy reward function, with a textual prompt containing a few examples or a description of the desired behavior. Lafite-RL \cite{chu2023accelerating} enables agents to learn robotic tasks more efficiently by leveraging timely feedback from LLMs.

Physical systems often involve diverse interconnected elements, necessitating the use of multiple agents to construct digital systems. These agents need to interact with each other, creating emergent behaviors such as competition and collaboration to simulate complex system-level phenomena. LLMs can facilitate more natural and flexible communication between agents through natural language, reducing the need for rigid communication protocols and enhancing explainability. For instance, Gao et al. \cite{gao2023s} proposed an LLM-based social network simulation system that models the propagation of information, attitudes, and emotions. CAMEL \cite{CAMEL} introduces a novel communicative agent framework that guides chat agents toward task completion while maintaining alignment with human intentions. ChatDev \cite{ChatDev} presents a chat-powered software development framework in which LLM-based agents contribute to the design, coding, and testing phases through unified language-based communication.


In summary, integrating LLMs into agent-based modeling offers several advantages that address the limitations of traditional methods. LLMs enhance agents by providing robust, flexible, and context-aware data analysis capabilities. LLM-based agents can exhibit reasoning and planning abilities comparable to symbolic agents \cite{xi2023rise} through techniques such as chain-of-thought and problem decomposition. They can also develop interactive capabilities with their environment, similar to reactive agents, by learning from feedback and adapting their actions. Additionally, LLMs can function as sophisticated world simulators, offering valuable feedback and rewards that drive the learning and development of agents. This integration ultimately supports more efficient and scalable simulation system modeling, overcoming the challenges of manual model design and extending applicability to diverse physical elements.

\subsection{LLM-enhanced Predictive Modeling}

Building upon reflective and experimentable digital systems, digital twins offer significant advantages in supporting predictive analysis and decision-making. However, existing digital twins often rely on the manual generation of virtual scenarios through parameter adjustments and the analysis of simulated data using human-crafted models to derive optimal control strategies. This approach is heavily dependent on expert knowledge and requires substantial manual effort, resulting in inefficiencies and limited scalability.

LLMs, demonstrating remarkable capabilities across various applications, have the potential to become valuable tools for designing and generating virtual scenarios. Leveraging their powerful natural language understanding and generation abilities, LLMs can engage in interactive conversations with humans to perform human-instructed scenario engineering tasks \cite{10423819}. Their planning and reasoning capabilities enable them to deconstruct complex tasks into more manageable subtasks, facilitating adaptive and effective task execution. Additionally, LLMs can generate data analysis code or utilize external tools to analyze generated artificial data, thereby simplifying the design of strategy learning models.

\subsubsection{LLM-enhanced experimental design} 

Existing methods for experimental design typically generate numerous virtual scenarios by dynamically adjusting key parameters, which are then classified based on specific factors \cite{10530101,10190119}. The desired scenarios for specified experimental goals are identified through this process. However, these methods are often inefficient and costly, as they require generating and evaluating a large number of scenarios.

Leveraging their powerful understanding, planning, and generation capabilities, LLMs offer an efficient paradigm for scenario engineering in computational analysis. First, LLMs demonstrate strong natural language interaction abilities, enabling them to comprehend design requirements expressed in natural language. This allows humans to design experiments directly using language, significantly lowering the barrier for applying predictive digital twin systems. Second, LLMs can reason, plan, and take proactive measures to achieve specific goals. They have shown emergent planning abilities, including goal reformulation \cite{xi2023self}, task decomposition, and plan adjustments in response to environmental changes. Consequently, LLMs are well-equipped to design virtual scenarios with complex goals and requirements.


Due to these capabilities, LLMs have been successfully applied in experimental design and scenario engineering. For example, SimpleTOD \cite{SimpleTOD} utilizes a causal language model trained on all sub-tasks, reformulated as a single sequence prediction problem for task-oriented dialogue, enhancing the model's understanding and execution of user-defined tasks. Guan et al. \cite{guan2023leveraging} proposed a novel paradigm that constructs an explicit world model and uses it for model-based task planning. LLM-Planner \cite{LLM-planner} employs LLMs as planners to follow natural language instructions and complete complex tasks in visually perceived environments. Li et al. \cite{10423819} introduced an LLM-based scenario engineering framework that defines trajectory scenarios and employs prompt engineering to generate complex and challenging scenarios for trajectory prediction. Du et al. \cite{FactoryDecoder} developed an LLM-incorporated tool for non-expert users in 3D engineering, enabling them to generate and modify digital twins using natural language inputs. Thus, LLMs provide an efficient and intelligent experimental design paradigm for predictive analysis, facilitating the generation of virtual scenarios through human language instructions.

\subsubsection{LLM-enhanced code generation} 

Previous studies have implemented virtual scenarios based on predefined templates or human-crafted models, which lack flexibility and make it difficult to integrate supplementary modules. Moreover, LLM-generated virtual scenarios may not align well with predefined simulation models, necessitating the construction of generative models to create more flexible scenarios.

Given that code data is included in their pretraining corpus, LLMs have demonstrated exceptional performance in completing and synthesizing code from natural language descriptions \cite{3534862}, sparking discussions about their potential to replace human programmers. Some LLMs have been specifically trained for programming tasks \cite{nijkamp2022codegen}, offering stronger support for the code generation of artificial scenarios. Furthermore, LLMs can generate explanations for the code they produce, improving its readability and comprehensibility. Qian et al. \cite{ChatDev} proposed an LLM-based framework for software development that enhances the quality of developed software, leading to better completeness, executability, and alignment with requirements. LLMs have also shown promising results in automatically designing neural network models. For instance, Wang et al. \cite{wang2023graph} introduced a GPT-4-based graph neural architecture search method that uses novel prompt classes to guide GPT-4 in generating graph neural architectures.

Software testing is critical for ensuring the quality and executability of generated models, and recent studies have successfully utilized LLMs for the automated generation of unit tests \cite{3663839}, significantly improving the quality of generated code. Leveraging these techniques, virtual scenarios can be automatically generated with the help of LLMs, greatly enhancing the efficiency of predictive analysis implementation.

\subsubsection{LLM-enhanced data analysis} 

Data analysis and visualization are essential for improving users' understanding of computational analysis results. Traditional digital twins typically rely on human-crafted tools to analyze and visualize key indicators, which often lack flexibility and require considerable human effort. Moreover, the human-machine interaction methods in these systems are not sufficiently intelligent, making it challenging for users to access the specific results they are interested in.

LLMs not only excel at generating data analysis code but are also effectively applied in utilizing external tools to achieve data visualization from language queries. Furthermore, they can be prompted to create or call data analysis tools to generate visualizations. For example, Maddigan \cite{10121440} proposed a novel system that leverages LLMs to convert free-form natural language into code for generating appropriate visualizations. Li et al. \cite{li2024prompt4vis} introduced a framework that enhances the performance of generating data visualizations from natural language. The framework includes two key components: a multi-objective mining module designed to identify effective examples that strengthen the LLM's in-context learning capabilities for text-to-visualization tasks, and a schema filtering module to simplify the database schema. With the support of LLMs, users can obtain the desired data analysis and visualization results through natural language queries, greatly enhancing flexibility and accessibility.

\subsubsection{LLM-enhanced strategy optimization} 

The goal of predictive analysis is to derive optimal strategies for specific scenarios, which requires the optimization of control strategies. Previous studies have proposed various learning methods to derive optimized control strategies from generated data \cite{10555241}. Despite their excellent performance, these methods often suffer from limited interactivity and explainability. Additionally, many optimization algorithms are derivative-based, which face significant challenges when gradients are absent.

LLMs have demonstrated exceptional pattern-matching and generalization capabilities in both natural language and other domains, positioning them as a promising tool for optimization tasks. Cheng et al. \cite{10757666} combined vision-based LLMs with digital twin frameworks to enable dynamic, real-time insights into urban safety, thereby fostering smarter and safer cities. Sun et al. \cite{SUN202483} introduced an LLM-driven multi-agent architecture that derives insights into digital twin systems through specific interaction mechanisms while ensuring traceability. Yang et al. \cite{yang2023large} proposed a simple yet effective approach where LLMs act as optimizers, generating new solutions from prompts containing previously generated solutions and their evaluations, and integrating these solutions into the optimization process. Liu et al. \cite{liu2024large} reformulated the Bayesian optimization problem in natural language, enabling LLMs to iteratively propose and evaluate promising solutions based on historical data. This approach has shown strong empirical performance in hyperparameter tuning. Similarly, LICO \cite{nguyen2024lico} employs LLMs as surrogate models for black-box optimization, incorporating a separate embedding layer and prediction layer to train the model to perform in-context predictions across a diverse set of functions. LLMs have been shown to perform comparably to or even outperform traditional hyperparameter optimization methods, such as Bayesian optimization, on standard benchmarks \cite{zhang2023using}. Furthermore, LLMs effectively tackle black-box optimization problems while providing interpretable solutions through step-by-step reasoning, enhancing the overall transparency and explainability of the optimization process.

\subsection{LLM-enhanced Prescriptive Modeling}

Prescriptive control in digital twins is rarely discussed, with only a few methods proposed, mainly due to the varying control interfaces and control flows of different physical systems. Most existing methods rely on human-crafted interfaces to transform learned strategies into machine-executable instructions for implementing optimal strategies within physical systems. Some approaches use predefined functions to recommend optimal strategies to managers for human-in-the-loop control \cite{9788029,9005386}. However, these methods require significant human effort to customize control interfaces for different applications. Additionally, their human-machine interaction interfaces are often not user-friendly, making their strategy recommendations difficult to interpret and implement.

Tool learning unlocks the potential of LLMs to effectively interact with a wide range of tools, including APIs and system interfaces, to perform complex tasks. By integrating LLMs with the APIs and interfaces of physical systems, LLMs can function as intelligent controllers, executing complex control strategies autonomously. Thanks to their long-context learning abilities, LLMs can learn to interact with APIs and interfaces directly from their documentation, without requiring extensive additional training. This capability significantly reduces human effort. Furthermore, LLMs can explain their control strategies and actions in natural language, thereby improving the interactivity and explainability of prescriptive control systems. This makes the decision-making process more transparent and accessible, enhancing the overall user experience in managing complex physical systems.

\begin{figure}[!htp]
  \centering
  \includegraphics[width=1\linewidth]{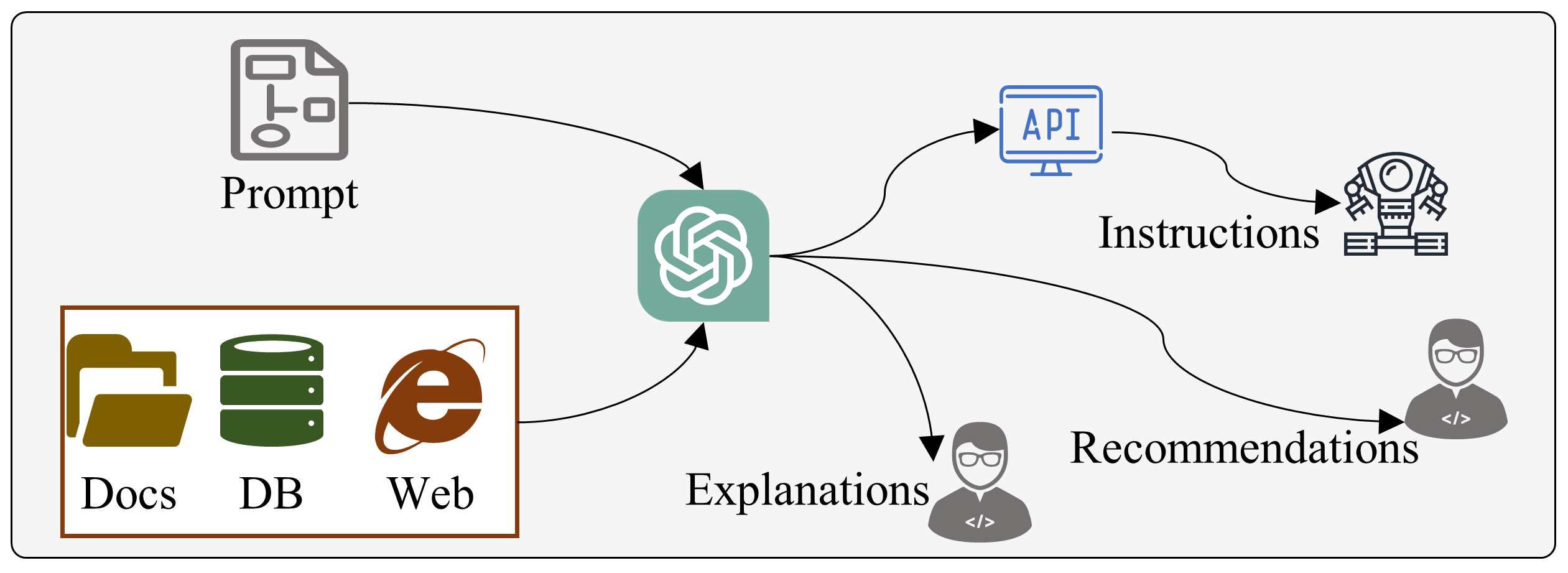}
  \caption{LLM-enhanced prescriptive modeling.}
\label{fig:LLM4P}
\end{figure}

\subsubsection{LLM-enhanced generative control} 

Existing digital twin systems often rely on manually designed functions to translate learned strategies into machine-executable or human-understandable instructions. For example, Jin et al. \cite{9788029,9005386} proposed various recommendation approaches to optimize signal timing parameters for urban traffic management and control. Similarly, Xu et al. \cite{10394535} introduced a parallel reasoning framework that generates power grid dispatching strategies from an artificial knowledge graph, enabling human-machine collaboration in managing complex power grids. However, these methods require substantial human effort to design specialized functions tailored to specific physical systems and tasks, which results in limited flexibility and labor-intensive processes.

With their robust natural language generation capabilities, LLMs offer significant potential for controlling physical systems through natural language interfaces. For instance, LLMs can rapidly generate accurate factual content to address public misconceptions during instances of widespread misinformation \cite{FACTGEN}. Leveraging their extensive world knowledge and advanced tool-manipulation capabilities, LLMs can execute grounded actions and interact with diverse real-world devices. Liang et al. \cite{10160591}, for instance, demonstrated how LLMs can translate natural language commands into robot policy code, defining functions or feedback loops to process perception outputs and parameterize control APIs. Similarly, GPT4Tools \cite{yang2024gpt4tools} uses low-rank adaptation to train an LLM on an instruction dataset, enabling the LLM to utilize various tools effectively. Furthermore, LLMs exhibit strong generalization capabilities for unseen APIs, adapting efficiently with only the provided API documentation \cite{qin2023toolllm}. This capability facilitates the efficient customization of LLM-based controllers based on the documentation of physical system control interfaces.

In addition to their generative strengths, LLMs have demonstrated effectiveness in zero-shot ranking for recommender systems, excelling in conditional ranking tasks and even rivaling traditional recommendation models when multiple candidate generators are employed \cite{1lm4recomend}. Their advanced conversational abilities also enable the creation of interactive recommendation interfaces between users and systems. For instance, Jan et al. \cite{10569919} employed an LLM agent to develop intuitive conversational interfaces, allowing users to query and interact with digital twins naturally. These advancements highlight the significant potential of LLMs to serve as controllers for more intelligent, flexible, and interactive management of physical systems.

\subsubsection{LLM-enhanced strategy explanation} 

Explainability of control strategies and learning processes is a critical requirement for many complex systems to ensure their security and reliability. To address this need, various explainable AI technologies have been proposed and applied to clarify the rationale behind recommended strategies. For example, Zhang et al. \cite{9506997} introduced a backpropagation-based deep explainer that leverages Shapley additive explanations to provide interpretable models for deep reinforcement learning-based emergency control applications. Despite these advancements, challenges persist in managing complex decision models and improving the interactivity of these explanations.

\begin{figure}[!htp]
  \centering
  \includegraphics[width=1\linewidth]{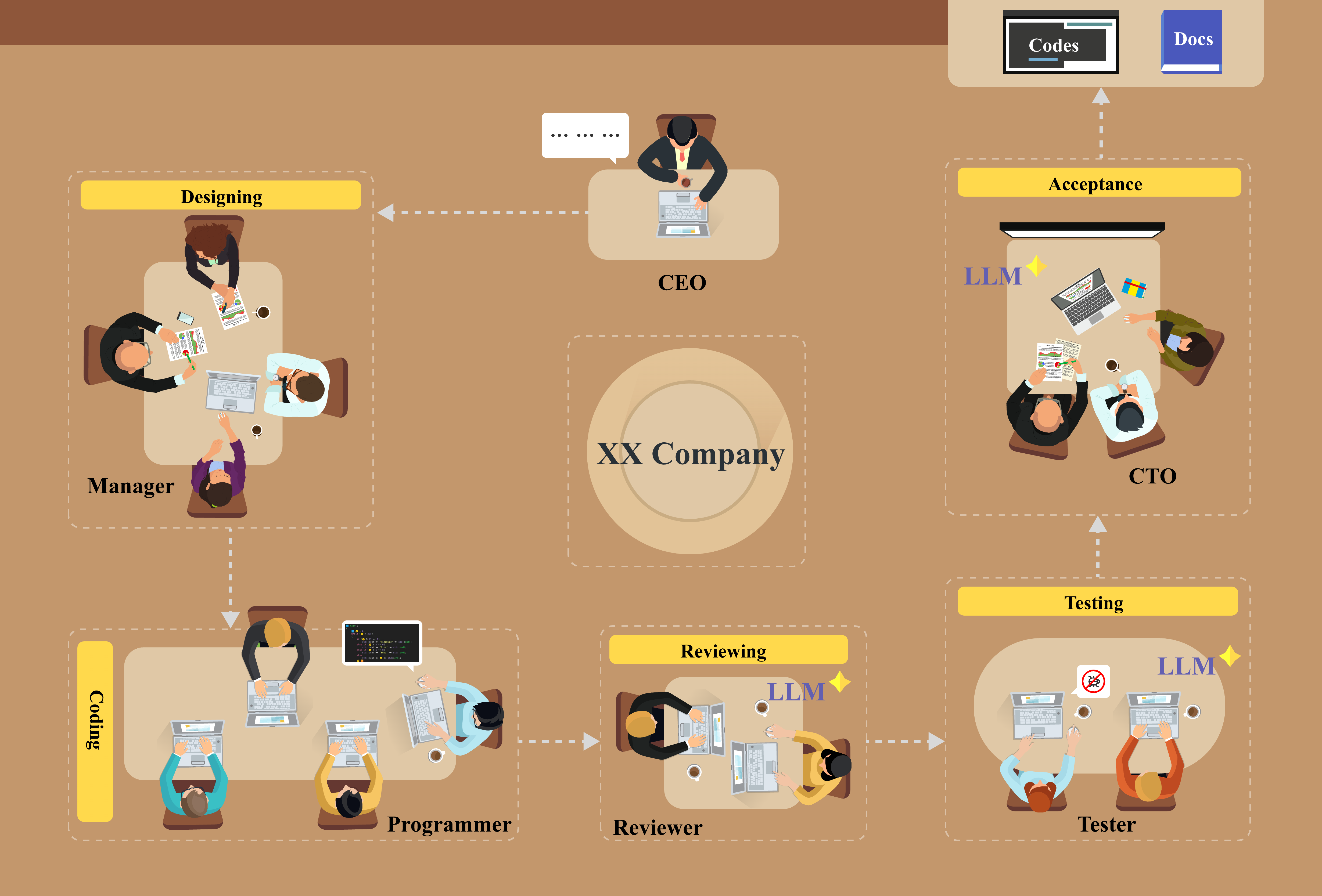}
  \caption{Illustration of the LLM-based digital enterprise system.}
\label{fig:case1}
\end{figure}

Recent research has focused on enhancing LLMs' ability to exhibit reasoning processes or explanations during the inference stage, with the aim of ensuring trustworthy predictions. For instance, Rajani et al. \cite{rajani-etal-2019-explain} developed a dataset with human annotations for commonsense reasoning, which trains LLMs to automatically generate explanations for their reasoning. Sammani et al. \cite{sammani2022nlx} introduced a compact and faithful language model capable of simultaneously predicting answers and providing explanations for vision and vision-language tasks. The reasoning and step-by-step explanation capabilities of LLMs are essential for supporting explainable digital twins. By emulating human-like thought processes, LLMs can generate intermediate reasoning steps that connect input data to final predictions. This approach not only enhances the interpretability of the model’s decisions but also offers a more detailed understanding of how conclusions are derived, helping to identify and correct potential biases or errors. Additionally, integrating external knowledge bases can further strengthen the credibility of LLMs' reasoning \cite{10417790}. For example, Zhang et al. \cite{zhang2024large} explored the use of LLMs to provide an explainability platform for dynamic, data-driven digital twins, generating natural language explanations of system decision-making by leveraging domain-specific knowledge bases.

In summary, LLMs can be effectively employed to explain generated strategies, contributing to more transparent and trustworthy control in complex systems.

\section{Case Study}

In this section, we demonstrate the effectiveness of LLMs in enhancing digital twins by developing an LLM-powered enterprise digital twin system. This system simulates the collaborative software development process within a technology company and optimizes the workflow for a specific software development project through computational analysis. This system aims to guide real-world companies toward improved software functionality and increased software development efficiency in their projects.

\subsection{System Design}

Software development is a multifaceted task that requires collaboration among various members and teams, typically involving three key phases: design, coding, and testing. Although numerous studies have utilized deep learning models to enhance specific phases of software development \cite{3505243}, these approaches often remain isolated due to technical inconsistencies. Recently, LLM-based agents have been integrated into various roles to create comprehensive solutions for software development \cite{ChatDev, hong2023metagpt}, enabling the simulation of software development processes. Building on these insights, we develop a chat-powered digital enterprise system to simulate the software development process of a real technology company. In this system, specialized LLM-driven agents collaborate to complete assigned software development tasks through natural language communication.

The architecture of the developed digital enterprise system is illustrated in Fig.~\ref{fig:case1}. The system employs five types of agents—CEO, manager, programmer, reviewer, and tester—each created by prompting an LLM. These agents are responsible for a variety of tasks, including proposing software project requirements, analyzing these requirements, coding, debugging, and testing. To enhance the realism of role simulations, daily records from each agent’s activities can be used to fine-tune the LLM or provide contextual information. In this study, we follow the methodology outlined in \cite{ChatDev} to prompt the LLM to assume different roles. For each subtask, two agents with distinct roles collaborate using natural language to complete the task, with one acting as the instructor and the other as the assistant. For example, in the code review subtask, one agent assumes the role of the code reviewer, providing feedback after reviewing the code. The other agent, acting as the programmer, refines the code based on the reviewer's feedback. Through iterative interactions between the two agents, the code is debugged and improved. Subtasks are executed sequentially according to a predefined workflow, with the output of one task serving as the input for the subsequent task, ensuring continuity and coherence throughout the process.

\begin{figure*}
    \centering
    \includegraphics[width=0.85\linewidth]{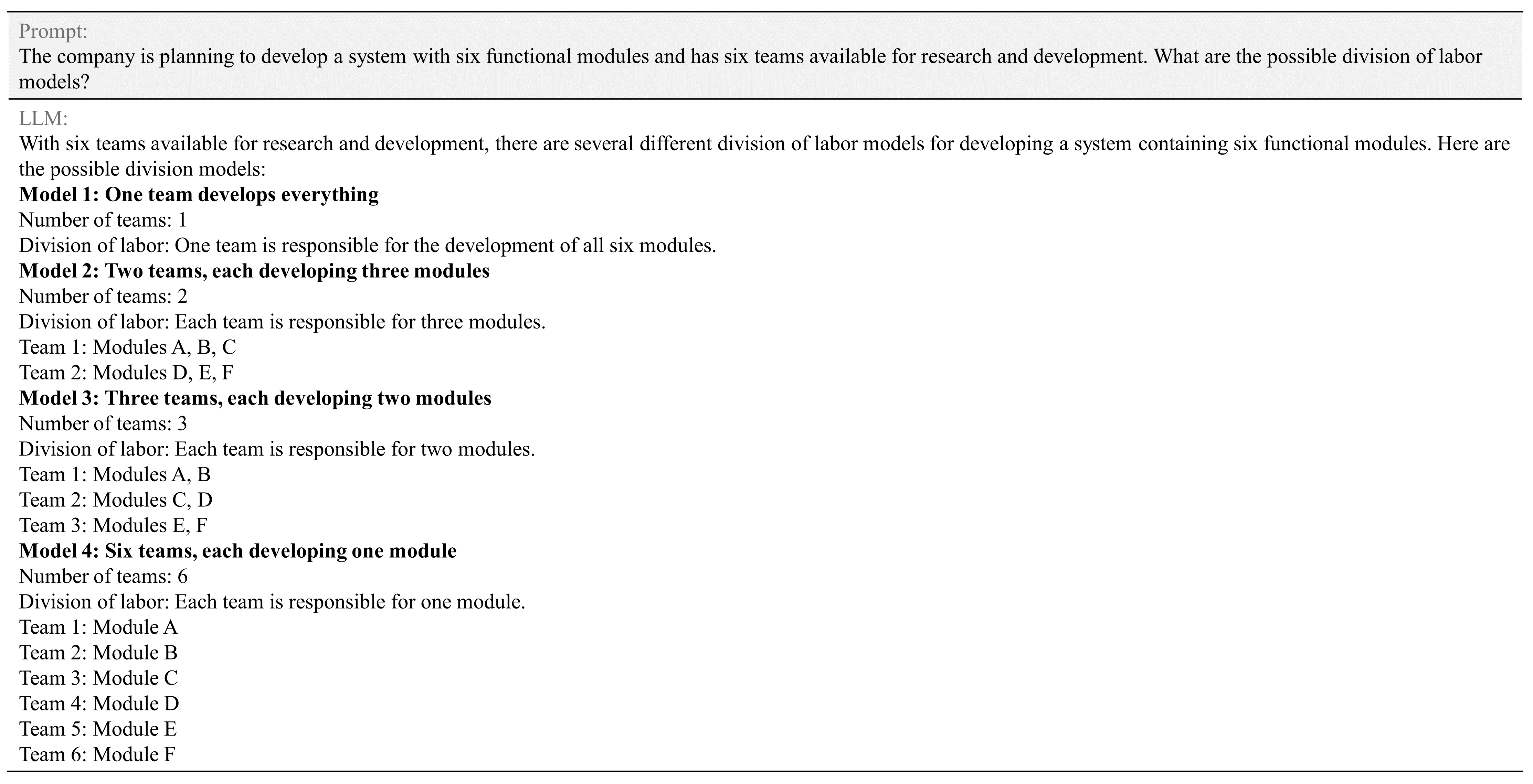}
    \caption{Computational experiment plan generation with LLM.}
    \label{fig:case2}
\end{figure*}

In this case study, we evaluate the effectiveness of the developed enterprise digital twin system by using the development of an enterprise data collection and analysis system as an example. This project involves creating software capable of collecting, processing, and analyzing various business data, which includes six functional modules: data collection, cleaning, storage, analysis, visualization, and report generation. With up to six software development teams involved in the project, an optimal project division plan must be generated, considering both cost and development efficiency.

To achieve this, we first prompt an LLM to generate candidate computational experiment designs. The prompt and the LLM's response are shown in Fig.~\ref{fig:case2}. The LLM's response demonstrates its ability to comprehend the experimental design instructions and generate corresponding experiment plans with accuracy. Notably, the LLM takes into account the balance of development tasks among different teams when designing the experimental plans, thereby reducing the number of experimental scenarios and avoiding the unnecessary consumption of computing resources on unrealistic scenarios. Following this process, the LLM-based digital enterprise system automatically executes the designed virtual scenarios, collecting the necessary data on development time and the number of software functions achieved. These metrics are then analyzed, and the optimal project division scheme is recommended to the company based on its cost and computational efficiency requirements.

\subsection{Result Analysis}

\begin{figure}
    \centering
    \includegraphics[width=0.91\linewidth]{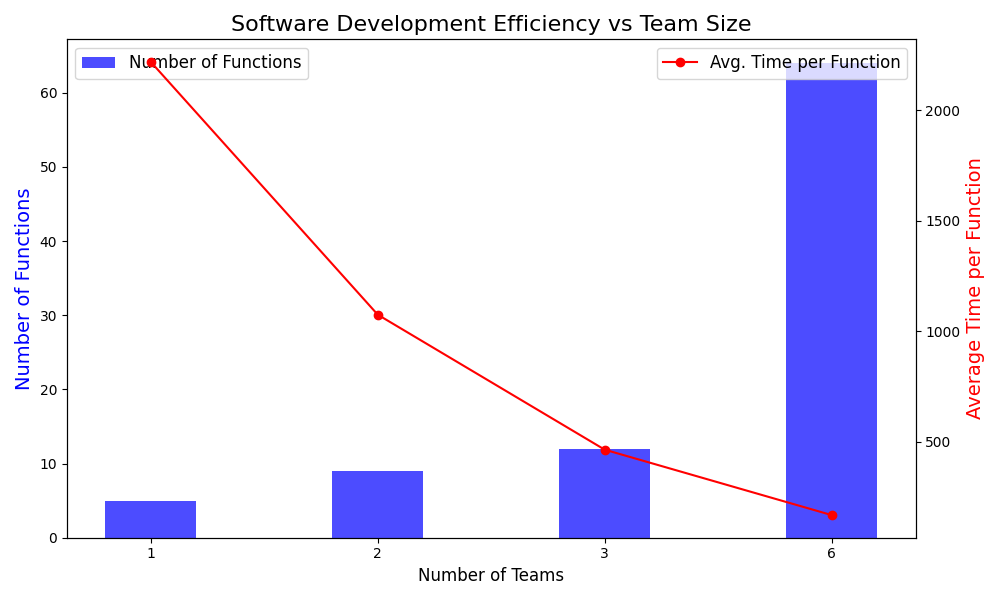}
    \caption{Computational experimental results.}
    \label{fig:case3}
\end{figure}

We implement the digital enterprise system and perform computational analysis using GPT-4 with a temperature setting of 0.2. As shown in Fig.\ref{fig:case2}, the LLM proposes four experimental schemes, dividing the software development project into 6, 3, 2, and 1 team(s) respectively. The number of functions developed and the corresponding software development efficiency are presented in Fig.\ref{fig:case3}. In this context, software development efficiency is defined as the average simulation time required to develop a function, with a lower value indicating higher efficiency. The results indicate that as the number of teams increases, both the number of functions developed and software development efficiency improve. This suggests that increased labor input leads to better outcomes and reduces the development time per function. Therefore, when labor costs are not considered, utilizing 6 teams to complete the project appears to be the optimal solution.

Based on the computational experiment results, the LLM can be prompted to select and recommend the most suitable software development scheme according to the company's specific needs. For example, if the company has sufficient personnel and aims to complete the project as quickly as possible, the LLM would recommend dividing the project into 6 teams. Additionally, the LLM can provide natural language explanations for its recommendations, thereby enhancing the trustworthiness of its suggestions. Thus, LLM-enhanced digital twins not only improve the intelligence and efficiency of the system but also provide greater explainability, offering a novel approach to addressing the challenges faced by traditional digital twins.

\section{Discussion}

The preceding sections have provided a comprehensive review of existing methods for enhancing the modeling of digital twins using LLMs. In this section, we discuss the challenges associated with these methods and propose future research directions, focusing on the current limitations of LLMs.

\subsection{Challenges for LLM-enhaced Digital Twins}

Although LLMs hold significant potential within the description-prediction-prescription framework, the success of LLM-enhanced digital twin systems is inherently tied to the capabilities and limitations of the underlying LLM. This section highlights the primary challenges faced by such systems, including their dependence on LLM capabilities, difficulties in handling heterogeneous data, inefficiencies in interaction, challenges in model updates, and broader ethical, legal, and safety concerns \cite{xi2023rise}.

\textit{Dependence on LLM capabilities.} The performance of LLM-enhanced digital twins is highly contingent on the capabilities of the underlying LLM. Different LLMs exhibit varying strengths across tasks, with some excelling in natural language processing while others perform better in coding or reasoning tasks \cite{3534862}. These differences stem from variations in their pretraining and fine-tuning corpora. Despite the extensive knowledge embedded in LLMs, limitations such as hallucinations, insufficient reasoning abilities, and knowledge obsolescence \cite{10417790} persist. These issues can lead to incorrect predictions and, in some cases, severe system failures. Therefore, integrating LLMs into digital twins requires rigorous evaluation to mitigate these risks and ensure reliability in real-world scenarios.

\textit{Heterogeneous data.} Digital twins depend on integrating diverse data types—ranging from structured tabular data to unstructured text, images, and sensor outputs—to accurately replicate and predict the behavior of physical systems. The fusion of such heterogeneous data poses significant challenges, particularly when the data modalities differ widely. Although LLMs demonstrate strengths in processing multi-source data, aligning and integrating disparate inputs remains a complex task. Furthermore, inconsistencies in the data, such as missing values, conflicting records, or outdated information, can undermine predictive accuracy and system reliability. While LLMs exhibit robustness to certain types of noise, persistent inconsistencies can amplify error propagation during inference. Addressing these issues requires robust preprocessing pipelines to harmonize disparate data sources and advanced fine-tuning techniques to adapt LLMs to specific domain requirements.

\textit{Model upates.} The dynamic nature of digital twins necessitates frequent updates to both the underlying data and models. Digital twins operate in environments where real-time changes in physical systems must be accurately reflected in their virtual counterparts. This demands continuous ingestion of new data and frequent model updates to maintain fidelity. However, LLMs are computationally intensive, requiring significant resources for retraining or fine-tuning. Frequent updates can introduce latency, increase operational costs, and complicate real-time synchronization. Addressing these challenges calls for the development of efficient incremental update methods, optimization of resource allocation, and approaches that maintain system responsiveness while ensuring accurate real-time performance.

\textit{Inefficient interactions.} LLMs exhibit several inefficiencies in interactive scenarios. A key challenge lies in the substantial computational resources required for real-time interactions, as their large parameter sizes lead to latency in response generation. Another issue is the verbosity of LLM outputs, where overly detailed responses can overwhelm users and detract from clarity and utility. Moreover, LLMs often struggle to maintain contextual consistency over extended interactions, necessitating repeated input of contextual information and reducing the fluidity of communication. Overcoming these inefficiencies will require advancements in algorithms to streamline response generation and improvements in hardware for faster inference.

\textit{Ethical, legal, and safety issues.} The deployment of LLMs in practical applications raises complex ethical, legal, and safety concerns. A major ethical issue involves the potential for LLMs to generate biased or harmful content. Ensuring fairness and minimizing bias require meticulous data curation and ongoing monitoring of model outputs, which can be resource-intensive and challenging to implement. From a legal perspective, LLMs often necessitate access to large volumes of personal data for training and fine-tuning, raising concerns around data privacy and intellectual property. Additionally, the unpredictability of LLM behavior and the potential for adversarial attacks pose significant safety risks, particularly in high-stakes domains such as healthcare, finance, and autonomous systems. Addressing these concerns demands stricter regulations, robust system safeguards, and the adoption of transparency measures to build trust in LLM-driven digital twin systems.

\subsection{Future Research Directions}

To address the technical challenges inherent in LLM-enhanced digital twins, we outline several promising research directions. While advancing the foundational capabilities of LLMs is crucial for the success of LLM-enhanced digital twins, such efforts fall outside the scope of this paper. Instead, we focus on practical strategies to overcome current limitations.

\textit{Developing general toolkits.} The application of LLMs in digital twin systems is still in its infancy, with a significant lack of platforms designed to seamlessly integrate LLMs into the description-prediction-prescription paradigm. This gap hinders the widespread adoption and advancement of LLM-enhanced digital twins. Developing specialized toolkits and frameworks is essential to enable efficient and effective integration of LLMs into the digital twin modeling pipeline. Such toolkits would facilitate robust connectivity between LLMs and key components, including databases, APIs, and tools, streamlining the development of digital twin systems. By addressing this need, these frameworks would unlock the full potential of LLMs in practical applications, significantly enhancing the efficiency and functionality of LLM-enhanced digital twins.

\textit{Exploring heterogenous data fusion methods.} Digital twins operate on diverse data sources, ranging from sensor outputs and textual data to simulation results and external knowledge bases. Effective fusion of such heterogeneous data is critical for accurately modeling and predicting the behavior of complex systems. Future research should focus on developing advanced data fusion techniques to enable LLMs to integrate multimodal inputs seamlessly. These methods would improve the accuracy and responsiveness of digital twins, enhancing their ability to handle the complexities of real-world applications.

\textit{Developing efficient model and data update mechanisms.} Digital twins rely on continuous streams of real-time data to mirror changes in the physical systems they represent. Consequently, LLMs supporting these systems must adapt dynamically to new information. However, the computational cost of frequently updating large models presents a major scalability challenge. To address this, future research should focus on creating efficient mechanisms for incremental model updates, dynamic data processing, and real-time synchronization. These innovations will ensure that LLM-enhanced digital twins remain up-to-date and adaptive, maintaining their performance and scalability in rapidly changing environments.

\textit{Incorporating external knowledge with LLMs.} Real-world applications often depend on domain-specific processes, regulations, and proprietary knowledge stored in local databases, which general-purpose LLMs—trained primarily on open web data—typically lack. To bridge this gap, retrieval-augmented generation (RAG) techniques can be employed to dynamically retrieve and integrate relevant information from private or domain-specific knowledge bases during training and inference. This approach not only enriches the contextual knowledge of LLMs but also mitigates issues such as hallucination, thereby improving the safety, accuracy, and reliability of their outputs. By enabling LLMs to incorporate specialized external knowledge, RAG techniques significantly expand their applicability in domain-specific digital twin systems \cite{10518077}.

\textit{Enhancing operational efficiency.} The computational overhead and latency associated with integrating LLMs into digital twins pose significant barriers to their practical adoption. Addressing these challenges requires optimizing both the algorithms and hardware infrastructure to reduce resource consumption and improve response times. Techniques such as model pruning, quantization, and the use of specialized hardware can play a pivotal role in achieving these goals. By improving operating efficiency, these advancements enable faster, more responsive digital twins that can handle complex tasks with greater accuracy and lower cost. Ultimately, such optimizations will make LLM-enhanced digital twins more accessible and practical across a wider range of applications.

\section{Conclusion}

Large language models, with their pretrained knowledge and advanced capabilities such as zero-shot learning and in-context learning, present a promising solution for enhancing digital twins across various dimensions, including descriptive, predictive, and prescriptive modeling. In this paper, we systematically reviewed recent advancements in LLMs and categorized their roles within the description–prediction–prescription framework. For each category, we analyzed the relevant literature, outlined methodologies, discussed the challenges addressed, and provided insights into future research directions.

By enhancing tasks such as data preparation and processing, agent modeling, experimental design, results analysis, control, and explanation, LLM-enhanced digital twins offer innovative solutions to improve the intelligence, efficiency, and interpretability of corresponding systems. To further illustrate the potential of LLM-enhanced digital twins, we developed an enterprise digital twin system powered by an LLM. This system simulates collaborative software development scenarios and generates optimal strategies by considering factors such as effectiveness and efficiency. It assists enterprises in designing optimal development workflows for specific software tasks, showcasing the practical value and transformative potential of LLM-enhanced digital twins.

\section*{Acknowledgment}
This work was supported by National Natural Science Foundation of China under Grant 62306288.



\bibliographystyle{elsarticle-num} 
\bibliography{reference}


%
%
%
\end{document}